\documentstyle[amstex,12pt,epsf]{article}

\begin{document}

\begin{center}
{\Large {\bf Phenomenon of the time-reversal violating magnetic field
generation by a static electric field in a medium and vacuum.} }

{\Large {\bf Vladimir G.Baryshevsky} }

{\Large \bigskip }

{\it Nuclear Problems Research Institute, Bobruiskaya Str. 11, }

{\it Minsk 220080 Belarus. }

{\it Electronic address: bar@@inp.minsk.by }

{\it Tel: 00375-172-208481 }

{\it Fax: 00375-172-265124}{\Large \ }

\bigskip
\end{center}

P- and T-odd interactions cause mixing of opposite parity levels of atom
(molecule) that yields to the appearance of P- and T-odd terms of the atom
(molecule) polarizability \cite{PLA93}. This makes possible to observe
various optical phenomena, for example, photon polarization plane rotation
and circular dichroism in an optically homogeneous medium placed to an
electric field, polarization plane rotation (circular dichroism) phenomena
for photons moving in an electric (gravitational) field in vacuum \cite
{PLA99}.

The energy of atom (molecule) in external electromagnetic field includes the
term caused by the time reversal violating interactions \cite{PLA93}:

\begin{equation}
\Delta U=-\frac{1}{2}\beta _{S}^{T}\overrightarrow{E}\overrightarrow{H},
\label{1}
\end{equation}
where $\beta _{S}^{T}$ is the scalar T-noninvariant polarizability of atom
(molecule), $\overrightarrow{E}$\ \ is the external electric field, $%
\overrightarrow{H}$\ is the external magnetic field.

It's well known \cite{Landau} that when the external field frequency $\omega
\rightarrow 0$ the polarizabilities describe the processes of magnetization
of medium by a static magnetic field and electric polarization of a medium
by a static electric field

The energy of interaction of magnetic moment $\overrightarrow{\mu }$ with
magnetic field $\overrightarrow{H}$

\begin{equation}
W_{H}=-\overrightarrow{\mu }\overrightarrow{H}  \label{2}
\end{equation}
Comparison of (\ref{1}) and (\ref{2}) let one to conclude that the action of
stationary electric field on an atom (molecule) induces the magnetic moment
of atom

\begin{equation}
\overrightarrow{\mu }(\overrightarrow{E})=\frac{1}{2}\beta _{S}^{T} 
\overrightarrow{E}
\end{equation}
On the other hand, the energy of interaction of electric dipole moment $%
\overrightarrow{d}$ with electric field $\overrightarrow{E}$

\begin{equation}
W_{E}=-\overrightarrow{d}\overrightarrow{E}.  \label{4}
\end{equation}
As it follows from (\ref{1}) and (\ref{4}), magnetic field induces the
electric dipole moment of atom

\begin{equation}
\overrightarrow{d}(\overrightarrow{H})=\frac{1}{2}\beta _{S}^{T} 
\overrightarrow{H}
\end{equation}
As appears from the above, atom (molecule) being placed to static electric
field gets the induced magnetic moment which in its part produces magnetic
field. And similarly, if atom (molecule) is placed in the area of magnetic
field the induced electric dipole moment yields to the appearance of its
associated electric field.

Let us consider the simplest possible experiment. Suppose that homogeneous
isotropic matter (liquid or gas) is placed to the area occupied by an
electric field $\overrightarrow{E}$. From the above it follows that the time
reversal violation yields to the appearance of magnetic field $%
\overrightarrow{H}_{T}=4\pi \rho \overrightarrow{\mu }(\overrightarrow{E})$
parallel to $\overrightarrow{E}$\ in this area ($\rho $ is the number of
atoms (molecules) of matter per $cm^{3}$ ). And vice versa, the electric
field $\overrightarrow{E}_{T}=4\pi \rho \overrightarrow{d}(\overrightarrow{H}
)$ appears under matter placement to the area occupied by a magnetic field $%
\overrightarrow{H}$. Let us estimate the effect value. It is easy to do by $%
\beta _{S}^{T}$ evaluation. The general case explicit expression for
polarizabilities for time dependent fields were derived in \cite{PLA93} (see
eqs. (12)-(20) therein).

Briefly the calculation technique is as follows. Let us suppose that atom is
placed to the arbitrary periodic in time electric and magnetic fields. The
energy of interaction of an atom (molecule) with these fields has the
routine form

\begin{equation}
W=-\widehat{\overrightarrow{d}}\overrightarrow{E}-\widehat{\overrightarrow{
\mu }}\overrightarrow{H}+.....  \label{interaction energy}
\end{equation}
where $\widehat{\overrightarrow{d}}$ is the operator of atom electric dipole
moment and $\widehat{\overrightarrow{\mu }}$ is the operator of atom
magnetic dipole moment 
\begin{equation}
\overrightarrow{E}=\frac{1}{2}\left\{ \overrightarrow{E}_{0}\;e^{-i\omega
t}+ \overrightarrow{E}_{0}^{\ast }\;e^{i\omega t}\right\} ,\;\overrightarrow{
H}= \frac{1}{2}\left\{ \overrightarrow{H}_{0}\;e^{-i\omega t}+ 
\overrightarrow{H} _{0}^{\ast }\;e^{i\omega t}\right\}
\end{equation}
The Shr\"{o}dinger equation describing atom interaction with electromagnetic
field is as follows:

\begin{equation}
i\hbar \frac{\partial \psi (\xi ,t)}{\partial t}=[H_{A}(\xi )+W(\xi ,t)]\psi
(\xi ,t),
\end{equation}
where $H_{A}(\xi )$ is the atom Hamiltonian taking into account the weak
interaction of electrons with nucleus in the center of mass of the system, $%
\xi $ is the space and spin variable of electron and nucleus, $W$ is the
energy of interaction of atom with electromagnetic field of frequency $%
\omega $

\begin{eqnarray}
W &=&Ve^{-i\omega t}+V^{+}e^{i\omega t},  \label{W} \\
V &=&-\frac{1}{2}(\overrightarrow{d}\overrightarrow{E_{0}}+\overrightarrow{%
\mu }\overrightarrow{H_{0}}),V^{+}=-\frac{1}{2}(\overrightarrow{d}%
\overrightarrow{E_{0}}^{\ast }+\overrightarrow{\mu }\overrightarrow{H_{0}}%
^{\ast })  \nonumber
\end{eqnarray}
Let us perform the transformation $\psi =\exp (-{i}\frac{{H_{A}}}{{\hbar }}{t%
})\varphi $. Suppose $H_{A}\psi _{n}=E_{n}\psi _{n}$ ($E_{n}=E_{n}^{(0)}-%
\frac{1}{2}i\Gamma _{n}$, $E_{n}^{(0)}$ is the atom level energy, $\Gamma
_{n}$ is the atom level width), then $\varphi =\sum_{n}b_{n}(t)\psi _{n}$.
Therefore it follows from (\ref{W}) 
\begin{gather}
i\hbar \frac{\partial b_{n}(t)}{\partial t}=\sum_{f}\left\{ \langle
n|V|f\rangle \exp [i(E_{n}-E_{f}-\hbar \omega )t/\hbar ]\right. + \\
+\left. \langle n|V^{+}|f\rangle \exp [i(E_{n}-E_{f}+\hbar \omega )t/\hbar
]\right\} b_{f}(t),\;\langle \psi _{n}|\psi _{m}\rangle \ll 1.  \nonumber
\end{gather}
Suppose $b_{n0}$ be the ground state amplitude. Let us substitute the
amplitude $b_{f}$ describing the excited atom state into the equation for $%
b_{n0}$ and study this equation\ at time $t\gg \tau _{f}=\hbar /\Gamma
_{f}\;($or $\tau _{f}=\hbar /\Delta E);\;\Delta E=E_{f}^{(0)}-E_{n0}-\hbar
\omega $; $\Gamma _{f}$ $\gg $ $|\langle n|V|f\rangle |$ (or $\Delta E\gg $ $%
|\langle n|V|f\rangle |$). Therefore $b_{n0}$ is defined by equation 
\[
i\hbar \frac{\partial b_{n0}(t)}{\partial t}=\widehat{U}_{eff}\;b_{n0},{\
where} 
\]
\begin{equation}
\widehat{U}_{eff}=-\sum_{f}\left( \frac{\langle n_{0}|V|f\rangle \;\langle
f|V^{+}|n_{0}\rangle }{E_{f}-E_{n0}+\hbar \omega }+\frac{\langle
n_{0}|V^{+}|f\rangle \;\langle f|V|n_{0}\rangle }{E_{f}-E_{n0}-\hbar \omega }%
\right)  \label{U_eff}
\end{equation}
Substituting $V$ and $V^{+}$ into\ (\ref{U_eff}) one can obtain 
\begin{equation}
\widehat{U}_{eff}=-\frac{1}{2}\widehat{g}_{ik}^{E}E_{0i}E_{0k}^{\ast }-\frac{%
1}{2}\widehat{g}_{ik}^{H}H_{0i}H_{0k}^{\ast }-\frac{1}{2}\widehat{g}%
_{ik}^{EH}E_{0i}H_{0k}^{\ast }-\frac{1}{2}\widehat{g}%
_{ik}^{HE}H_{0i}E_{0k}^{\ast },
\end{equation}
where the polarizability of atom (molecule) is: 
\begin{eqnarray}
\widehat{g}_{ik}^{E} &=&-\frac{1}{2}\left( \sum_{f}\frac{\langle
n_{0}|d_{i}|f\rangle \;\langle f|d_{k}|n_{0}\rangle }{E_{f}-E_{n0}+\hbar
\omega }+\frac{\langle n_{0}|d_{k}|f\rangle \;\langle f|d_{i}|n_{0}\rangle }{%
E_{f}-E_{n0}-\hbar \omega }\right)  \nonumber \\
\widehat{g}_{ik}^{H} &=&-\frac{1}{2}\left( \sum_{f}\frac{\langle n_{0}|\mu
_{i}|f\rangle \;\langle f|\mu _{k}|n_{0}\rangle }{E_{f}-E_{n0}+\hbar \omega }%
+\frac{\langle n_{0}|\mu _{k}|f\rangle \;\langle f|\mu _{i}|n_{0}\rangle }{%
E_{f}-E_{n0}-\hbar \omega }\right)  \nonumber \\
\widehat{g}_{ik}^{EH} &=&-\frac{1}{2}\left( \sum_{f}\frac{\langle
n_{0}|d_{i}|f\rangle \;\langle f|\mu _{k}|n_{0}\rangle }{E_{f}-E_{n0}+\hbar
\omega }+\frac{\langle n_{0}|\mu _{k}|f\rangle \;\langle
f|d_{i}|n_{0}\rangle }{E_{f}-E_{n0}-\hbar \omega }\right)  \nonumber \\
\widehat{g}_{ik}^{HE} &=&-\frac{1}{2}\left( \sum_{f}\frac{\langle n_{0}|\mu
_{i}|f\rangle \;\langle f|d_{k}|n_{0}\rangle }{E_{f}-E_{n0}+\hbar \omega }+%
\frac{\langle n_{0}|d_{k}|f\rangle \;\langle f|\mu _{i}|n_{0}\rangle }{%
E_{f}-E_{n0}-\hbar \omega }\right)  \nonumber
\end{eqnarray}
It should be noted that $\widehat{g}_{ik}^{E}$ and $\widehat{g}_{ik}^{H}$
are the P- and T-invariant electric and magnetic polarizability tensors and $%
\widehat{g}_{ik}^{EH}$ and $\widehat{g}_{ik}^{HE}$ are the P- and
T-noninvariant polarizability tensors

Let an atom be placed at the static ($\omega \rightarrow 0$) magnetic and
electric fields $\overrightarrow{E}$ and $\overrightarrow{H}$ of the same
direction. Then it's perfectly easy to obtain the effective energy of P- and
T-odd interaction of an atom with these fields.

\begin{equation}
\widehat{U}_{eff}^{T,P}=-\frac{1}{2}\left( \sum_{f}\frac{\langle
n_{0}|d_{z}|f\rangle \;\langle f|\mu _{z}|n_{0}\rangle +\langle n_{0}|\mu
_{z}|f\rangle \;\langle f|d_{z}|n_{0}\rangle }{E_{f}-E_{n_{0}}}\right) EH
\end{equation}
Axis $z$ is supposed to be parallel to $\overrightarrow{E}$. Thus from (\ref
{1}) 
\begin{equation}
\beta _{S}^{T}=\sum_{f}\frac{\langle n_{0}|d_{z}|f\rangle \;\langle f|\mu
_{z}|n_{0}\rangle +\langle n_{0}|\mu _{z}|f\rangle \;\langle
f|d_{z}|n_{0}\rangle }{E_{f}-E_{n_{0}}}
\end{equation}
Let us estimate the $\beta _{S}^{T}$ order of magnitude. The atom state \ $%
|f\rangle $\ does not possess the certain parity because of weak T-odd
interactions. And over the weakness of $V_{T}$\ the state $|f\rangle $\ is
mixed with the state of opposite parity of value $\eta _{T}=\frac{{V_{W}^{T}}%
}{{E_{f}-E_{n}}}$. According to 
\begin{equation}
\beta _{S}^{T}\sim \frac{\langle d\rangle \;\langle \mu \rangle }{%
E_{f}-E_{n_{0}}}\eta _{T}
\end{equation}

For the heavy atoms the mixing coefficient can attain the value $\eta
_{T}\approx 10^{-14}$. Taking into account that matrix element $\langle \mu
\rangle \sim \alpha \langle d\rangle $ (where $\alpha =\frac{1}{137}$ is the
fine structure constant) one can obtain $\beta _{S}^{T}\sim \eta
_{T}\;\alpha \frac{{\langle d\rangle ^{2}}}{{\Delta }}{\;}\approx
10^{-16}\cdot \frac{{8\cdot 10^{-36}}}{{10^{-12}}}\approx 10^{-40}$.
Therefore, the electric field $E=10^{2}\;CGSE$ induces magnetic moment $\mu
_{T}\approx 10^{-38}$. Then, the magnetic field in the liquid target can be
estimated as follows

\begin{equation}
H=4\pi \rho \mu _{T}\approx 10^{23}\cdot 10^{-38}=10^{-15}\;gauss
\end{equation}
The magnitude of magnetic field strength can be increased, for example, by
tightening of the magnetic field with superconductive shield. In this way
the measured field strength can be increased by four orders when one collect
the field from the area 1 $m^{2}$ to the area 1 $cm^{2}$ (Fig.1).

\begin{figure}[h]
\epsfxsize = 10 cm \centerline{\epsfbox{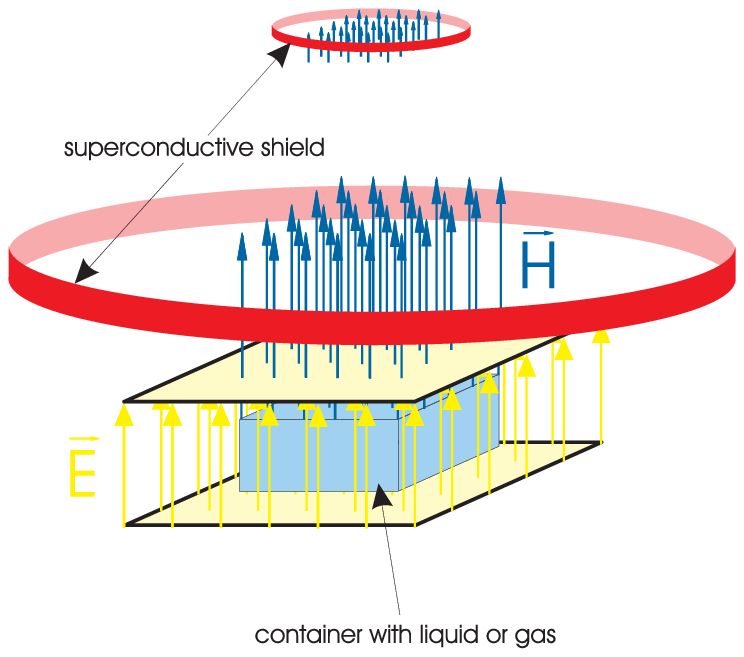}}
\caption{ }
\end{figure}

The induced magnetic moment produces magnetic field at the electron
(nucleus) of the atom. This field $H^{T}(E)\sim \mu \;\langle \frac{{1}}{{%
r^{3}}}\rangle \sim 10^{-38}\cdot 10^{26}=10^{-12}$ $gauss$. Therefore, the
frequency of precession of atom magnetic moment $\mu _{A}$ in the magnetic
field induced by an external electric field

\begin{equation}
\Omega _{E}\sim \frac{\mu _{A}\;\beta \;E\;\langle \frac{{1}}{{r^{3}}}%
\rangle }{\hbar }=\frac{10^{-20}\cdot 10^{-12}}{10^{-27}}=10^{-5}\;\sec ^{-1}
\end{equation}
It should be reminded that to measure the electric dipole moment the shift
of precession frequency of atom spin in the presence of both magnetic and
electric fields is investigated. Then, the T-odd shift of precession
frequency of atom spin includes two terms: frequency shift conditioned by
interaction of atom electric dipole moment with electric field $\omega _{E}=%
\frac{{2d_{A}E}}{{\hbar }}$ and frequency shift $\Omega =\frac{{2\mu H^{T}(E)%
}}{{\ \hbar }}$ defined above. This aspect should be considered when
interpreting the similar experiments. One should take note of the mixing
coefficient $\eta _{T}$ essential increase when the opposite parity levels
are close to each other or even degenerate. Then the effect can grow up as
much as several orders $10^{5}\div 10^{6}$ (this occurs, for example, for
Dy, TlF, BiS, HgF).

The similar phenomenon of magnetic field induction by electric field can
occur in vacuum too.

Due to quantum electrodynamic effect of electron-positron pair creation in
strong electric, magnetic or gravitational field, the vacuum is described by
the dielectric $\varepsilon _{ik}$ and magnetic $\mu _{ik}$\ permittivity
tensors depending on these fields. The theory of $\varepsilon _{ik}$ \cite
{Ahiezer} does not take into account the weak interaction of electron and
positron with each other. Considering the T- and P-odd weak interaction
between electron and positron in the process of pair creation in an electric
(magnetic, gravitational) field one can obtain the density of
electromagnetic energy of vacuum contains term $\beta _{v}^{T}( 
\overrightarrow{E}\overrightarrow{H)}$ similar (\ref{1}) (in the case of
vacuum polarization by a stationary gravitational field $\beta _{g}^{T}( 
\overrightarrow{H}\overrightarrow{n_{g}})$, $\overrightarrow{n_{g}}=\frac{%
\overrightarrow{g}}{g}$, $\overrightarrow{g}-$ gravitational acceleration).

As a result both electric and magnetic fields (directed along the electric
field) could exist around an electric charge. But in this case $\oint 
\overrightarrow{B}d\overrightarrow{S}\neq 0$ ($\overrightarrow{B}$\ is the
magnetic induction) that is impossible in the framework of classic
electrodynamics. The existence of such field would means the existence of
induced magnetic monopole. If the condition $\oint \overrightarrow{B}d%
\overrightarrow{S}=0$ is fulfilled then for the spherically symmetrical case
the field appears equal to zero. Surely, the value of this magnetic field is
extremely small, but the possibility of its existence is remarkable itself.

The above result can be obtained in the framework of general Lagrangian
formalism. Lagrangian density can depend only on the field invariants. Two
invariants are known for the quasistatic electromagnetic field: $( 
\overrightarrow{E}\overrightarrow{H})$ and $(E^{2}-H^{2})$. In conventional
T-invariant theory these invariants are included in the Lagrangian $L$ only
as $(E^{2}-H^{2})$ and $(\overrightarrow{E}\overrightarrow{H})^{2}$, i.e. $%
L=L(E^{2}-H^{2},(\overrightarrow{E}\overrightarrow{H})^{2})$ \cite{Ahiezer}.
But while taking into account the T-odd interactions the Lagrangian can
include \ invariant $(\overrightarrow{E}\overrightarrow{H})$ raising to the
odd power, i.e.

\begin{equation}
L_{T}=L_{T}(E^{2}-H^{2},(\overrightarrow{E}\overrightarrow{H})^{2},( 
\overrightarrow{E}\overrightarrow{H}))  \label{lagrangian}
\end{equation}
Expanding (\ref{lagrangian}) by weak interaction one can obtain

\begin{equation}
L_{T}=L(E^{2}-H^{2},(\overrightarrow{E}\overrightarrow{H})^{2})+\beta _{T}( 
\overrightarrow{E}\overrightarrow{H}),
\end{equation}
where $L$ is the density of Lagrangian in P- and T-invariant
electrodynamics, $\beta _{T}=\beta _{T}(E^{2}-H^{2},(\overrightarrow{E} 
\overrightarrow{H})^{2})$ \ is the constant \ can be found in certain
theory. The explicit form of $L$ is cited in \cite{Ahiezer}.

The additions caused by the vacuum polarization can be described by the
field dependent dielectric and magnetic permittivity of vacuum. According to 
\cite{Ahiezer} the electric induction vector $\overrightarrow{D}$ and
magnetic induction vector $\overrightarrow{B}$ are defined as:

\begin{equation}
\overrightarrow{D}=\frac{\partial L}{\partial \overrightarrow{E}},\; 
\overrightarrow{B}=-\frac{\partial L}{\partial \overrightarrow{H}}
\end{equation}
Similarly the electric polarization $\overrightarrow{P}$ and magnetization $%
\overrightarrow{M}$ of vacuum can be found \cite{Ahiezer}:

\begin{eqnarray}
\overrightarrow{P}=\frac{\partial (L_{T}-L_{0})}{\partial \overrightarrow{E}}
,\;\overrightarrow{M}=-\frac{\partial (L_{T}-L_{0})}{\partial 
\overrightarrow{H}}, \\
\overrightarrow{D}=\overrightarrow{E}+4\pi \overrightarrow{P},\; 
\overrightarrow{B}=\overrightarrow{H}+4\pi \overrightarrow{M.}
\end{eqnarray}
In accordance with the above, the T-noninvariance yields to the appearance
of additional P- and T-odd terms to the electric polarization $%
\overrightarrow{P}$ and magnetization $\overrightarrow{M}$ . There are the
addition to the vector of electric polarization $\overrightarrow{P}$
proportional to the magnetic field strength $\overrightarrow{H}$ and the
addition to the vector of magnetization $\overrightarrow{M}$ proportional to
the electric field strength $\overrightarrow{E}$

\end{document}